\documentclass[journal,twocolumn]{IEEEtran}
\usepackage{xcolor}
\usepackage{mathtools}
\usepackage{epsfig,makeidx,color,subfigure}
\usepackage{amsmath,amssymb,bbm,enumitem}
\usepackage{cite,graphicx,lipsum}
\usepackage[switch,pagewise]{lineno}
\usepackage{hyperref}
\hypersetup{
        colorlinks = true,
        citecolor=blue,
}
\usepackage{wrapfig}
\usepackage[demo]{tikzpeople}
\usepackage{draftfigure}
\usepackage{tcolorbox}
\newtcolorbox{myprompt}[1][]{
    left=1.5mm,
    right=1.5mm,
    top=1.5mm,
    bottom=1.5mm,
  colback=gray!20,
  colframe=black!75!black,
  boxrule=0.5pt,
  fonttitle=\bfseries,
  #1
}

\begin{document}

% \title{\fontsize{23}{26}\selectfont Towards Semantic Communication Protocols: From Protocol Learning to Language-Oriented Approaches}
\title{\fontsize{21}{26}\selectfont Towards Semantic Communication Protocols for 6G: From Protocol Learning to Language-Oriented Approaches}

\author{
Jihong Park, Seung-Woo Ko, Jinho Choi, Seong-Lyun Kim, and Mehdi Bennis\thanks{
J. Park and J. Choi are with the School of Information Technology, Deakin University. VIC 3220, Australia (email: \{jihong.park, \ jinho.choi\}@deakin.edu.au). 

S.-W. Ko is with the Department of Smart Mobility Engineering, Inha University, Incheon 21999, Korea (e-mail: swko@inha.ac.kr).

S.-L. Kim is with the School of Electrical and Electronic Engineering, Yonsei University, Seoul 03722, Korea (email: slkim@yonsei.ac.kr). 

M. Bennis is with the Centre for Wireless Communications, University of Oulu, Oulu 90014, Finland (email: mehdi.bennis@oulu.fi)..}}

\maketitle
\begin{abstract}
The forthcoming 6G systems are expected to address a wide range of non-stationary tasks. This poses challenges to traditional medium access control (MAC) protocols that are static and predefined. In response, data-driven MAC protocols have recently emerged, offering ability to tailor their signaling messages for specific tasks. This article presents a novel categorization of these data-driven MAC protocols into three levels: Level 1 MAC. task-oriented neural protocols constructed using multi-agent deep reinforcement learning (MADRL); Level 2 MAC. neural network-oriented symbolic protocols developed by converting Level 1 MAC outputs into explicit symbols; and Level 3 MAC. language-oriented semantic protocols harnessing large language models (LLMs) and generative models. With this categorization, we aim to explore the opportunities and challenges of each level by delving into their foundational techniques. Drawing from information theory and associated principles as well as selected case studies, this study provides insights into the trajectory of data-driven MAC protocols and sheds light on future research directions.
\end{abstract}

\begin{IEEEkeywords}
Semantic protocol, protocol learning, multi-agent deep reinforcement learning (MADRL), large language model (LLM), 6G.
\end{IEEEkeywords}

\section{Introduction} \label{Sec:Intro}
Akin to language, a communication protocol in the medium access control (MAC) layer is a set of grammar and rules that govern the composition and exchange of signaling messages among network nodes to carry out communication control actions. This parallel to human language is not accidental – just as a language enables humans to communicate and understand one another, a MAC protocol enables devices to communicate and coordinate with each other. However, while human language naturally evolves and adapts over time to changing needs and contexts, traditional MAC protocols are human-crafted via standardization, with predefined rules and signaling messages. This approach has been successful in the past, but is being seriously challenged with the looming advent of new, diverse service verticals in 6G \cite{saad20network}.

Compared to previous generations, 5G considers three distinct service types, i.e., enhanced mobile broadband (eMBB), massive machine-type communication (mMTC), and ultra-reliable and low-latency communication (URLLC). To accommodate these service types, significant effort has been exerted to tailor MAC protocols to satisfy the key performance indicators (KPIs) of each service \cite{anand2020joint}. Looking ahead, 6G is anticipated to encompass an even broader range of verticals, each with their task-sensitive requirements such as autonomous cars in vehicle-to-everything (V2X) networks and visuo-haptic immersive applications in the metaverse. Furthermore, 6G services are expected to address non-stationary environments, such as satellites in non-terrestrial networks (NTNs) as well as control-communication co-designed communication (CoCoCo) for remote-controlled drones, cars, and mobile robots \cite{saad20network,park2022extreme}. The challenge of optimizing and standardizing a new protocol for every conceivable mission-critical and industrial tasks is undeniably becoming unsustainable in 6G.

\begin{figure*}
    \centering
    \includegraphics[width=.75\textwidth]{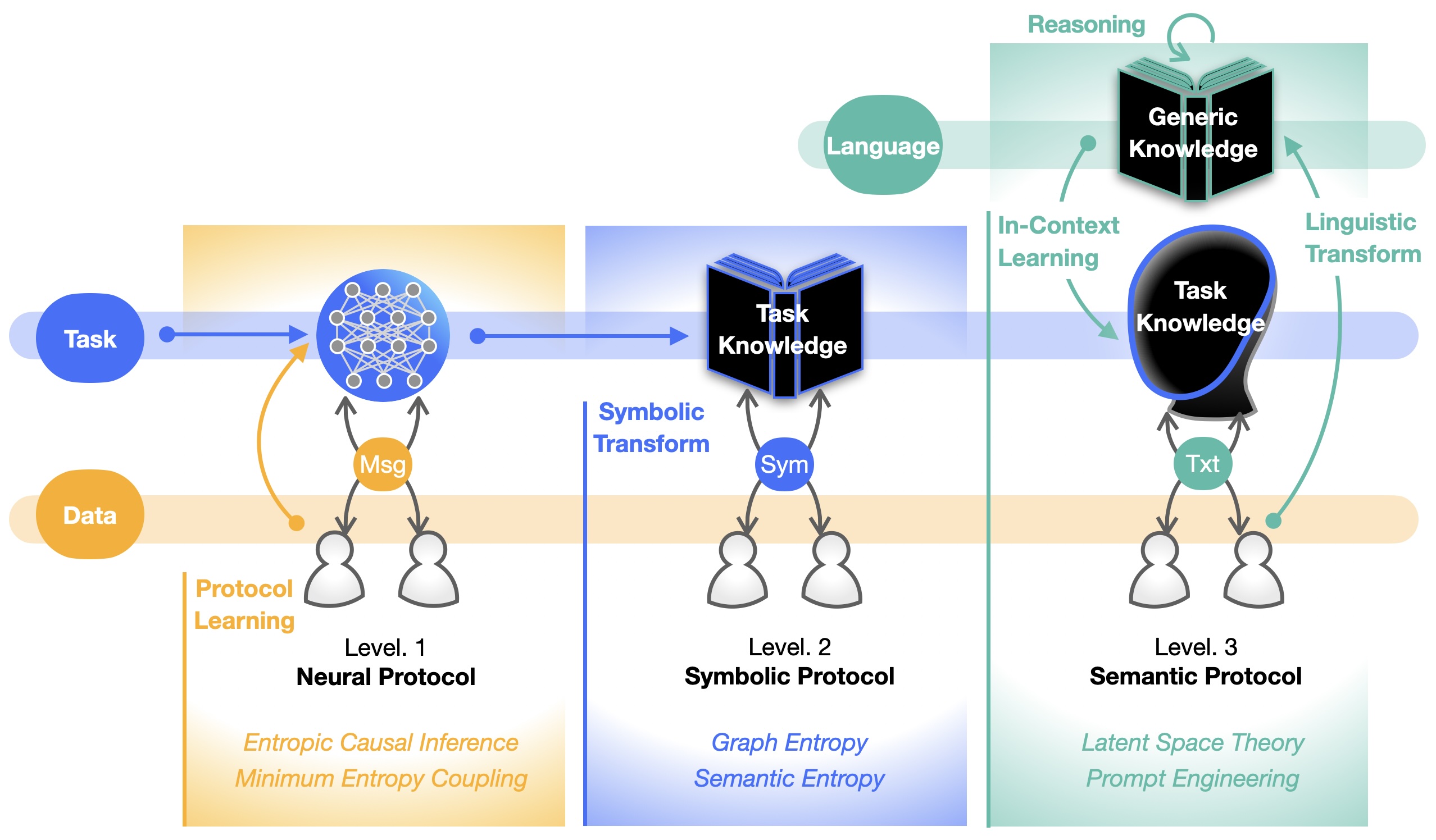}
    \caption{Three-level categorization of data-driven MAC protocols: Level 1. task-oriented neural protocols with task knowledge in black-box NNs; Level 2. NN-oriented symbolic protocols with symbolized task knowledge; and Level 3. language-oriented semantic protocols with both general and task-specific knowledge via large-language models (LLMs) and cross-modal generative models.}
    \label{fig:summary}
\end{figure*}

Departing from traditional task-agnostic MAC protocols, recently, there has been a great interest in learning MAC protocols directly from task-specific data and environments by embracing machine learning (ML) techniques \cite{valcarce21tabular,mota2021emergence,seo2022towards}. As Fig.~\ref{fig:summary} illustrates, these data-driven MAC protocols are inherently customized for given tasks, which can be categorized into three levels.

\textbf{Level 1 MAC - Task-Oriented Neural Protocols}.\quad
The current studies in data-driven MAC protocols are primarily centered on multi-agent deep reinforcement learning (MADRL) \cite{mota2021emergence}, which we classify under the \emph{Level 1 MAC} protocols. These Level 1 MAC protocols are \emph{task-oriented}, in the sense that signaling messages originate from inter-agent communication tailored to given tasks. Technically speaking, Level 1 MAC functions as \emph{neural protocols} where agents structured as neural networks (NNs) exchange their hidden-layer outputs as signaling messages during action executions within the MADRL framework. Though these message may initially appear arbitrary, they progressively acquire significance for inter-agent coordination as agents learn to refine actions and improve task efficiency. Consequently, bypassing expensive standardization processes, the emergent mappings from signaling messages to agents' states and control actions are innately suited for the task at hand.

\textbf{Level 2 MAC - NN-Oriented Symbolic Protocols}.\quad
While Level 1 MAC protocols successfully reflect task characteristics, they obscure task knowledge within their black-box NNs. Therefore, Level 1 MAC protocol operations are hardly interpretable, and any operational adjustments require NN re-training. Alternatively, \emph{Level 2 MAC} transforms neural protocols into \emph{symbolic protocols}, by identifying and transforming consistent messages from Level 1 MAC into explicit symbols~\cite{seo2022towards}. The resultant symbolic protocols present interpretable protocol operations, allowing direct manipulation. As Level 2 MAC protocols come from a neuro-symbolic transformation and thus \emph{NN-oriented}, advancements in Level 1 MAC can directly lead to improvements in Level 2 MAC, emphasizing the importance of optimizing Level 1 MAC design. However, feasible manipulations in Level 2 MAC are limited by the original task knowledge stored in the NN of Level 1 MAC. For adapting to non-stationary task environments, both Level 1 and Level 2 MAC protocols necessitate NN re-training.

\textbf{Level 3 MAC - Language-Oriented Semantic Protocols}.\quad
The advent of large language models (LLMs) hints at the potential for acquiring general knowledge in the domain of human language \cite{openai2023gpt4,touvron2023llama}. Concurrently, recent advances in cross-modal generative models, such as BLIP for image-to-text conversion \cite{li2022blip}, advocate the feasibility of translating various data modalities into human language. Inspired from these two recent developments, \emph{Level~3 MAC} protocols adopt a human \emph{language-oriented} approach, capturing general knowledge from different data types that can be transformed into human language. To address specific tasks, this general knowledge in Level~3 MAC can be fine-tuned to task-specific knowledge by providing contexts related to the tasks. Remarkably, LLMs can achieve this fine-tuning even without re-training, simply by demonstrating a few text-written examples, also known as in-context learning \cite{brown2020language}. As a result, Level~3 MAC introduces \emph{semantic protocols} that are inherently interpretable and semantically adjustable to cater to specific tasks. These semantic manipulations leverage both general and task knowledge, thereby facilitating reasoning beyond individual tasks and empowering flexible adaptations to non-stationary task environments.

To unlock the full potential of data-driven MAC protocols, this article aims to present their limitations and opportunities through the lens of information theory and other fundamental principles underlying Levels 1-3 MAC protocols, including entropic causal inference \cite{kocaoglu2017entropic}, semantic entropy \cite{Bao11}, and LLM latent space theory \cite{jiang2023latent}. Selected case studies will be showcased to corroborate the effectiveness of these principles and foster discussions on prospective research directions.

\section{Level 1 MAC: Task-Oriented Neural Protocol}

\subsection{Motivation}

\textbf{Task-Oriented and AI-Native Communication}.\quad
Emerging 6G applications are expected to have stringent latency and energy requirements, particularly in non-stationary and mission-critical environments. To address these demands, departing from the prevailing transmitter-centered and task-agnostic communication systems, there is a growing emphasis on receiver-centered and task-oriented communication frameworks. One notable example is pull-based communication, where transmissions are initiated solely based on receivers' requests \cite{chiariotti2022query}. The timing for these transmission opportunities can be determined by the age of outdated receiver information, also known as the age-of-incorrect information (AoII) \cite{maatouk2020age}. 

On the other hand, to cater to the challenges of unknown source data and channel distributions, AI-native communication frameworks are gaining traction. Within these frameworks, transceivers's end-to-end functionality is realized through an NN that ideally acts as a universal function approximator. These neural transceivers are trained directly with actual source and channel data while executing specific tasks, which aligns seamlessly with the task-oriented design paradigm. Recent AI-native communication frameworks have demonstrated their effectiveness in various tasks ranging from image and video reconstruction to recommendation and control, achieved by jointly optimizing source-channel coding, analog-digital modulation, and other pre/post-processing functionalities~\cite{gunduz2022beyond}.
% \cite{hornik1989multilayer}
% \cite{qin2022semantic}

However, much of the current research primarily concentrates on the physical layer (PHY) within a point-to-point setting. With slight modifications, the advantages and core principles of task-oriented and AI-naitve communication frameworks can be extended to higher layers, notably the MAC layers. Building on this perspective and viewing a MAC protocol as a language emerged from specific tasks, we will delve into task-oriented language emergence through MADRL as elaborated next.

\textbf{MADRL-based Emergent Communication}.\quad
Reinforcement learning (RL) seeks to determine a long-term optimal action $a\in\mathcal{A}$ for a given observed state $o\in\mathcal{O}$, or equivalently the policy $p(a|o)$. Deep RL aims to achieve this goal by rewarding action decisions and learning policies using NNs. While standalone deep RL considers a single decision-maker, i.e. a single actor or a single agent, MADRL involves multiple agents in a shared environment. Therefore, it is instrumental in coordinating different agents for cooperative, competitive, or hybrid tasks. To coordinate agents, a prevailing method is rewarding by a centralized entity during training, i.e., a centralized critic, resulting in the centralized-training and decentralized execution (CTDE) framework. 
% \cite{mnih2016asynchronous}
While effective in simple environments, decentralized execution may not always be functional under complicated inter-agent interactions, such as interference and collision. 
The MARL with communication frameworks \cite{zhu2022survey} aim to address this challenge, by adding a cheap-talk communication exchanging small control signaling messages across different agents for better inter-user coordination. 

It is however non-trivial to manually design such cheap-talk communication messages for MADRL. Alternatively, the MARL with communication framework is a prominent direction, wherein agents exchange their NN's hidden-layer outputs across agents in both training and execution. These NN outputs are random values at the beginning, but they gradually become meaningful as the training progresses. After training completes, the NN outputs can be regarded as emergent neural messages for coordinating multiple agents. 
The first of its kind is the differentiable inter-agent learning (DIAL) and reinforced inter-agent learning (RIAL) algorithms that consider continuous and discrete communication channels, respectively \cite{foerster2016learning}. Both RIAL and DIAL rely on sequential communication wherein each agent takes an action and sends a neural message to the next agent one by one. Extending this sequential operations to a parallel architecture, it is possible to exploit the same principle in protocol learning as we will explore in the next section.

\subsection{Key Technique: Neural Protocol Learning}

Protocol learning facilitates Level 1 MAC by building upon the MADRL with a communication framework. This section outlines the step-by-step extension from the standard MADRL to protocol learning.

\begin{figure}
    \centering
    \includegraphics[width=\columnwidth]{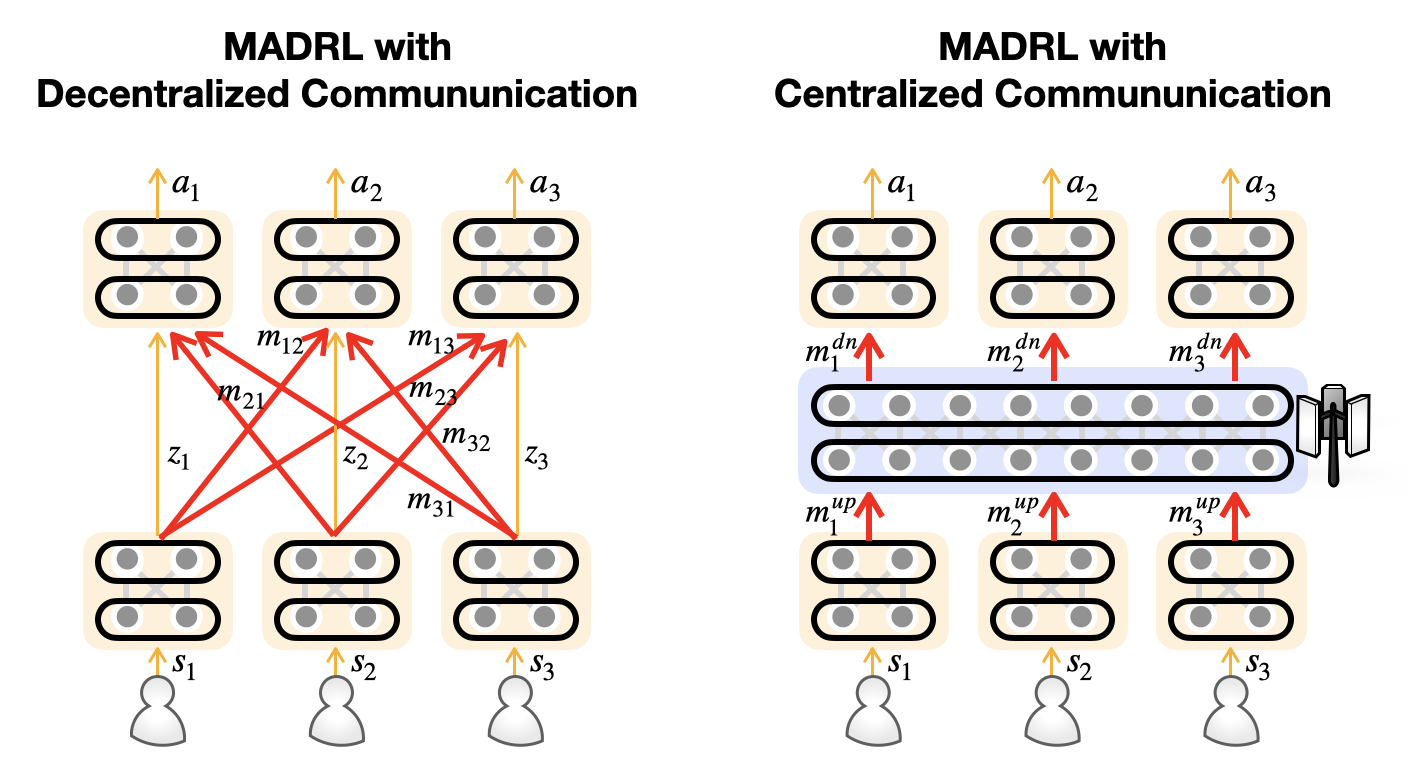}
    \caption{MADRL with decentralized communication (left) and centralized communication, i.e., protocol learning (right).}
    \label{fig:MADRL_architecture}
\end{figure}

\textbf{MADRL without Communication}.\quad
Consider an actor-critic MADRL architecture with a set $\mathcal{J}$ of agents. The $j$-th agent's decision-making is carried out by training the actor NN $\theta_j$, a function that returns an action $a_j$ for a given state $s_j$, which is represented as $a_j = \theta_j( s_j)$. We divide the actor NN into its upper segment $\theta^u_j$ and lower segment $\theta^l_j$:
\begin{align}
\theta_j = \begin{bmatrix} \theta^u_j \\ \theta^l_j \end{bmatrix}.
\end{align}
Let $z_j = \theta_j^l(s_j)$ denote the lower segment's output.
Accordingly, the functional relation is given as: 
\begin{align}
a_j = \theta_j^u( z_j ).
\end{align}
This actor NN architecture follows CTDE \cite{mnih2016asynchronous} wherein agents can only interact with others during training, but during execution, as illustrated in Fig.~\ref{fig:MADRL_architecture}.

\textbf{MADRL with Decentralized Communication}.\quad
Following RIAL and DIAL \cite{foerster2016learning}, the previous actor NN structure can be extended to an actor NN with cheap-talk communication as follows. At first, consider that the lower segment $\theta^l_j$ now generates two types of outputs: 1) a latent embedding $z_j $ that is the input of the upper segment $\theta^u_j$; and 2) an uplink control signaling message $m_{jk}$ from the $j$-th agent to the $k$-th agent, which are given as 
\begin{align}
\Big( z_j , z_j^\text{up} \Big) = \theta_j^l(s_j), \label{Eq:MADRL_up}
\end{align}
where $z_j^\text{up}=\bigoplus_{k\in\mathcal{J}\backslash j} m_{jk}$ and $\bigoplus$ implies the vector concatenation operation.
Likewise, the $j$-th agent receives a set of downlink control signal messages from other agents, i.e., $z_j^{\text{dn}}=\bigoplus_{k\in\mathcal{J}\backslash j} m_{kj} $. These received signaling messages are fed into the upper segment $\theta^u_j$, resulting in the final action as follows:
\begin{align}
a_j= \theta_j^u \Big (  z_j,  z_j^\text{dn}\Big ). \label{Eq:MADRL_dn}
\end{align}
During the early training phase, the signaling messages $\{z_j^\text{up},z_j^\text{dn}\}$ are meaningless random values. As the training progresses, these messages become gradually useful. Finally, after the training completion when all the actor NNs are capable of producing optimal actions $\{a_j\}$, the emergent messages $\{z_j^\text{up},z_j^\text{dn}\}$ are effective in inter-user coodrination. This cheap-talk extended actor NN architecture is visualized in Fig.~\ref{fig:MADRL_architecture}. Here, since each agent communicates directly with all users in a decentralized way, the communication costs in both uplink and downlink, i.e., $|z_j^\text{up}|$ and $|z_j^\text{dn}|$ increase with the number of agents.

\textbf{Protocol Learning - MADRL with Centralized Communication}.\quad
Protocol learning extends the previous MADRL with decentralized communication framework while making it scalable by reducing the uplink/downlink communication costs. To this end, protocol learning considers a centralized entity, a base station (BS). Precisely, consider a BS equipped with an NN $\theta^m$ that is shared by all agents and located in between $\theta_j^u$ and $\theta_j^l$, i.e., 
\begin{align}
\theta_j = \begin{bmatrix} \theta^u_j \\ \theta^m \\\theta^l_j \end{bmatrix}.
\end{align}
Then, the $j$-th agent uploads the signaling message $m_j^\text{up}$ to the BS and downloads the message $m_j^\text{dn}$ produced by the BS:
\begin{align}
m_j^\text{dn}= \theta^m \Big( \bigoplus_{k\in\mathcal{J}} m_j^\text{up} \Big). 
\end{align}
Applying $z_j^\text{up}=m_j^\text{up}$ and $z_j^\text{dn}=m_j^\text{dn}$ to \eqref{Eq:MADRL_up} and \eqref{Eq:MADRL_dn} finalizes the centralized communication for protocol learning. The MADRL architecture can be implemented in various ways. In \cite{valcarce21tabular}, a tabular Q-learning is applied. To cope with many agents by avoiding high gradient variance during training, multi-agent deep deterministic policy gradient (MADDPG) is applied in \cite{mota2021emergence}. To address dynamic inter-agent interactions, multi-actor-attention-critic (MAAC) is applied in \cite{shiri21maac}, which leverages Transformer NNs with the attention mechanisms.

\begin{figure}
    \centering
    \includegraphics[width=.85\columnwidth]{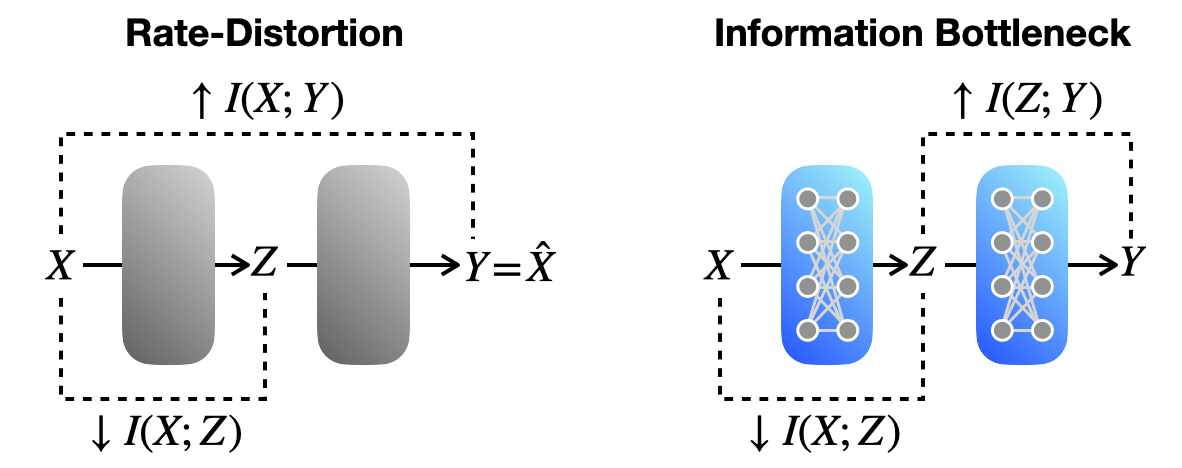}
    \caption{Rate-distortion (left) and information bottleneck (right) princples.}
    \label{fig:RD_IB}
\end{figure}

\subsection{Key Principles: Entropic Causal Inference and Minimum Entropy Coupling} \label{Sec:Lv1principle}

Information theory provides a useful insight for learning minimal signaling messages while maintaining the message effectiveness. To illustrate this with an example, consider a single-cell MAC protocol learning scenario with two users, Alice and Bob. Alice follows a single-cyle operation $S\overset{{\text{UL}}}{\to} Z \overset{{\text{BS}}}{\to} Y \overset{{\text{DL}}}{\to} A$, where Alice observes its state $S$, and sends an uplink signaling message $Z$ to a BS. The BS produces a downlink message $Y$ that makes the user take an access action $A$. Likewise, Bob's single-cycle operation follows $S'\overset{{\text{UL}}}{\to} Z' \overset{{\text{BS}}}{\to} Y \overset{{\text{DL}}}{\to} A'$, in which $Y$ is the same for both Alice and Bob. Focusing on $Z$, we aim to minimize $H(Z)$ while effectively running the protocol operations. Obviously, too small $H(Z)$ cannnot yield effective $Y$ as well as $A$ and $A'$, calling for balancing $H(Z)$.

If $Y=S$, as Fig.~\ref{fig:RD_IB} visualizes, we can apply classical rate-distortion (R-D) optimization, i.e., $\min_{p(z|s)} I(Z;S)$ with respect to the conditional distribution $p(z|s)$, subject to a distortion constraint or equivalently increasing $I(S;Y)$. If $Y\neq S$, information bottleneck principle (IB) \cite{tishby2015deep} provides an alternative solution to obtain minimal-yet-effective $Z$ by solving the following optimization:
\begin{align}
\max_{p(z|s)} I(Z;Y) - \beta I(Z;S), \label{Eq:IB}
\end{align}
for a constant $\beta \geq 1$. As visualized in Fig.~X, while both R-D and IB methods decrease $I(Z;S)$, R-D increases $I(S;Y)$, which is replaced with increasing $I(Z;Y)$ in IB. However, both R-D and IB methods can at most identify dependency, and fail to distinguish causes and their effects. For example, a causal relation $Z\to Y$ and its anti-causal relation $Y\to Z$ have the same mutual information $I(Z;Y)=I(Y;Z)$. The resultant spurious relations impede learning effective messages in tasks.

\begin{figure}
    \centering
    \includegraphics[width=.8\columnwidth]{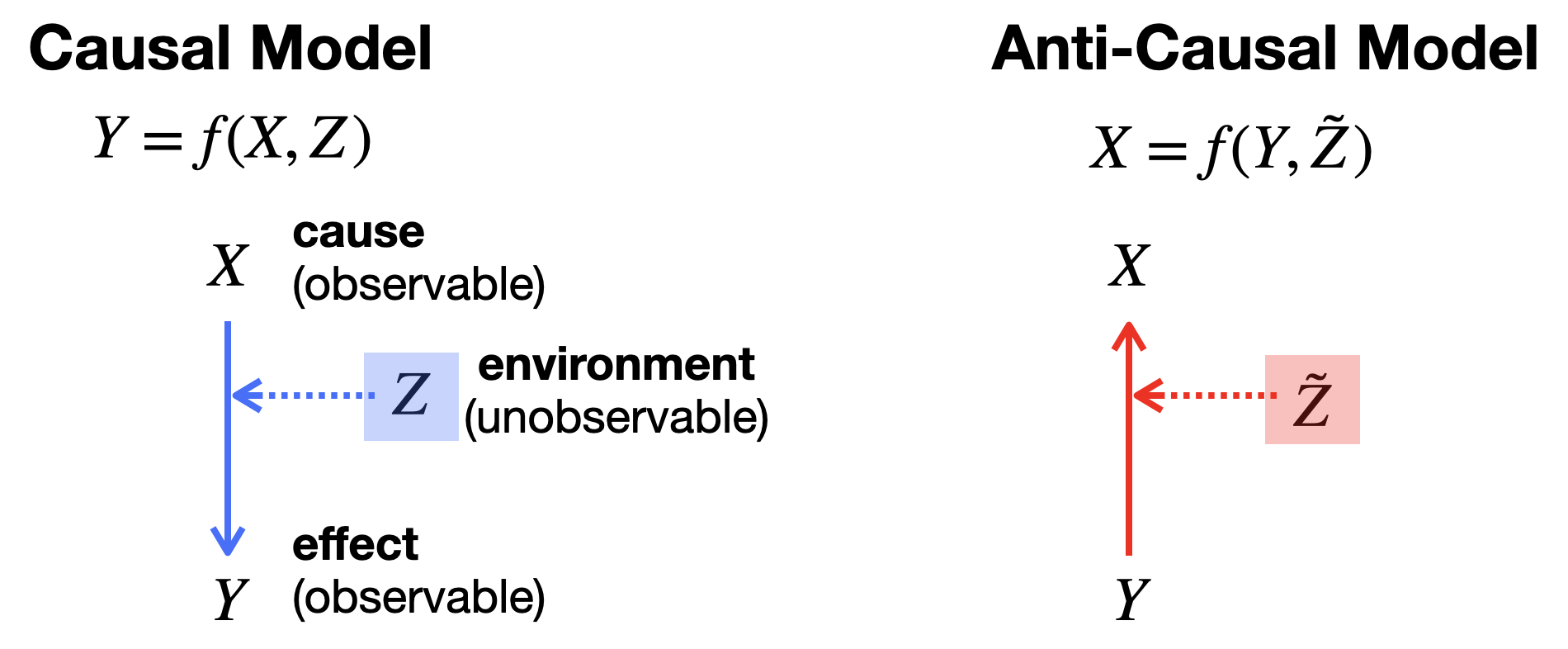}
    \caption{Causal (left) and anti-causal models (right) in entropic causal inference.}
    \label{Fig:causal}
\end{figure}

Alternatively, the framework of entropic causal inference~\cite{kocaoglu2017entropic} is promising to capture causality using $H(Z)$.
To be specific, for simplicity, we hereafter let $X=Z'$, and focus on Bob. Recall that Bob's uplink $X$ causes downlink $Y$ under Alice's uplink $Z$ that is unobservable from Bob, i.e., $Z\perp\!\!\!\!\perp X$. This causal relation can be written as $Y=f(X,Z)$ for a function $f(\cdot)$ such as a set of neural network layers. Next, as visualized in Fig.~\ref{Fig:causal}, consider an anti-causal relation $X = f'(Y,\tilde{Z})$ for some $\tilde Z \perp\!\!\!\!\perp X$. To distinguish the truth causal relation from its false anti-causal relation, entropic causal inference relies on an experimental phenomenon: $H(X) + H(Z) < H(Y) + H(\tilde{Z})$. For a given observable cause $X$ and its effect $Y$, this implies that the entropy $H(Z)$ of the true unobservable is lower than the entropy $H(\tilde{Z})$ of any false unobservable. According to \cite{kocaoglu2017entropic}, the best lower bound on $H(Z)$ is known as $H(Z) \geq H(U_1, U_2, \cdots, U_n)$ where $U_i:=(Y|X=i)$. Consequently, the minimal $H(Z)$ that maintains given causal relations or equivalently conditional observations $\{U_i\}$ can be derived by solving the following constrained joint probability minimization problem:
\begin{align}
\inf H(Z) = &\min_{p(u_1, u_2, \cdots, u_n)} H(U_1, U_2, \cdots, U_n) \nonumber\\
&\text{s.t.} \sum_{u_j, j\neq i} p(u_1, u_2, \cdots, u_n) = p(u_i), \forall i. \label{Eq:ECI}
\end{align}
This problem coincides with the minimum entropy coupling problem that can be solved by a polynomial-time algorithm within a $1$-bit gap \cite{cicalese2019minimum}.

\subsection{Case Study: Causality-Boosted Neural Protocol Learning} \label{Sec:Lv1Case}
\begin{figure}
    \centering
    \includegraphics[width=.85\columnwidth]{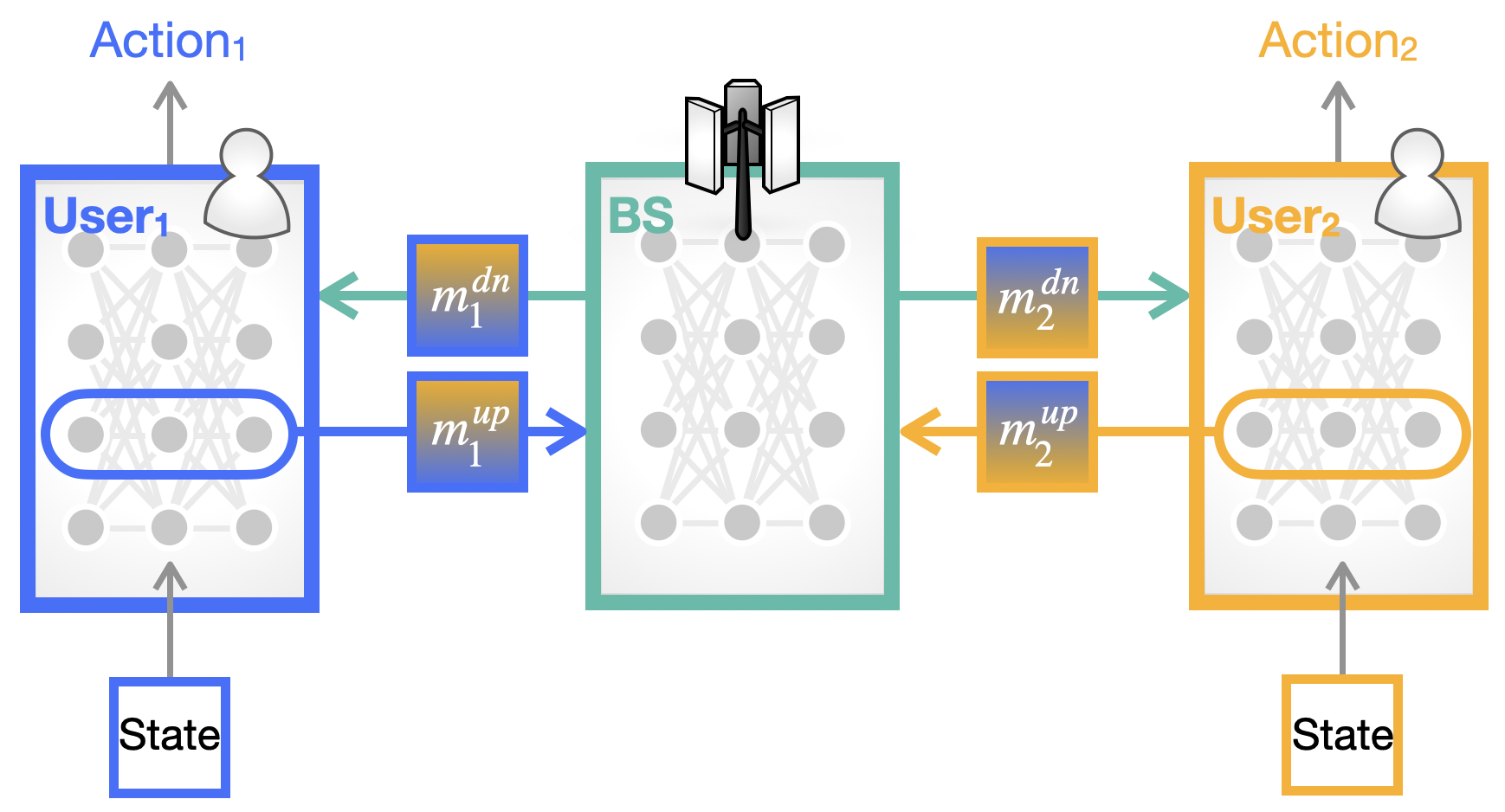}
    \caption{Level 1. Neural protocol: A single-cell network architecture example with two users.}
    \label{Fig:Lv1architecture}
\end{figure}

The operations of a neural protocol includes several causal relations which are implicitly facilitated during its training. Identifying key causation and boosting it can improve not only the effectiveness of the neural protocol in Level 2 MAC, but also the compactness of the neuro-symbolic semantic protocol in Level 3 MAC. To illustrate this, we revisit the single-cell MAC protocol learning scenario in Sec.~\ref{Sec:Lv1principle} with a set $\mathcal{J}$ of two users, as visualized in Fig.~\ref{Fig:Lv1architecture}. Following the settings in \cite{seo2022towards,mota2021emergence}, the $j$-th user equipment (UE) having its buffer as the {state} $s_j\in\{0,1,2\}$ sends an {uplink message} $m^\text{up}_j$ to the BS, and downloads a {downlink message} $m^\text{dn}_j$ from the BS, followed by taking an {action} $a_j$ out of possible actions: BS access (A), silence (S), and discarding the oldest buffer data (D). These state, uplink message, downlink message, and action correspond to an actor NN's input, a lower hidden layer's output, an upper hidden layer's output, and the final layer's output, respectively. Given these two actor NNs, during training, a shared critic NN gives a reward $\rho>0$ or $-\rho$ upon the buffer data reception success or failure, respectively.

\begin{figure*}
    \centering
    \subfigure[Average reward.]{\includegraphics[width=.6\columnwidth]{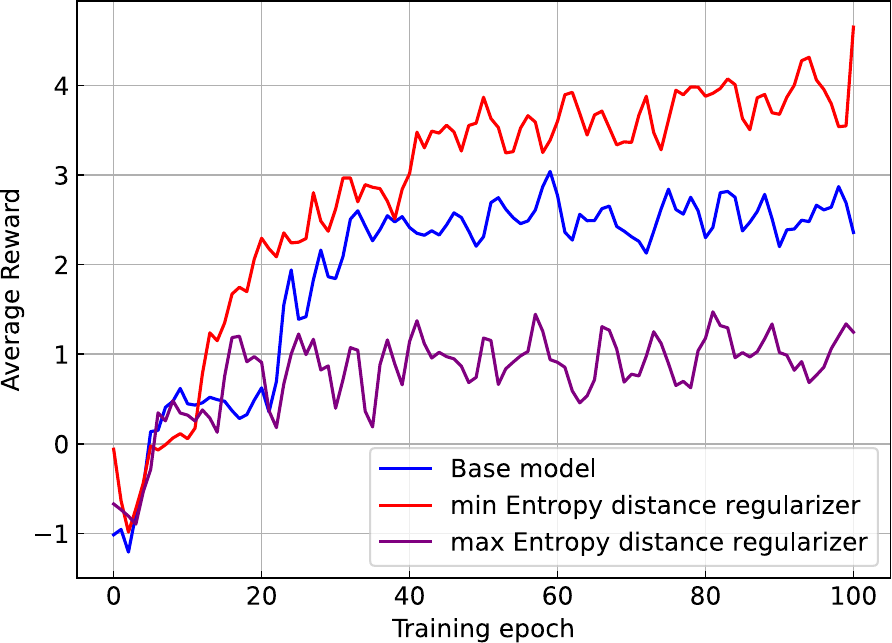}}    \hspace{10pt}
    \subfigure[Entropy.]{\includegraphics[width=.6\columnwidth]{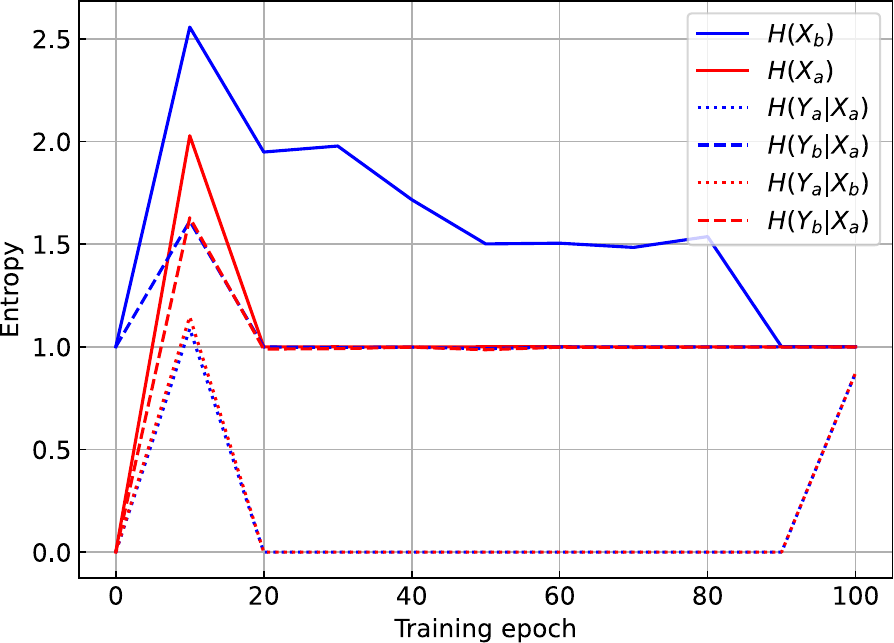}}
    \hspace{10pt}
    \subfigure[Codeword sparsity.]{\includegraphics[width=.6\columnwidth]{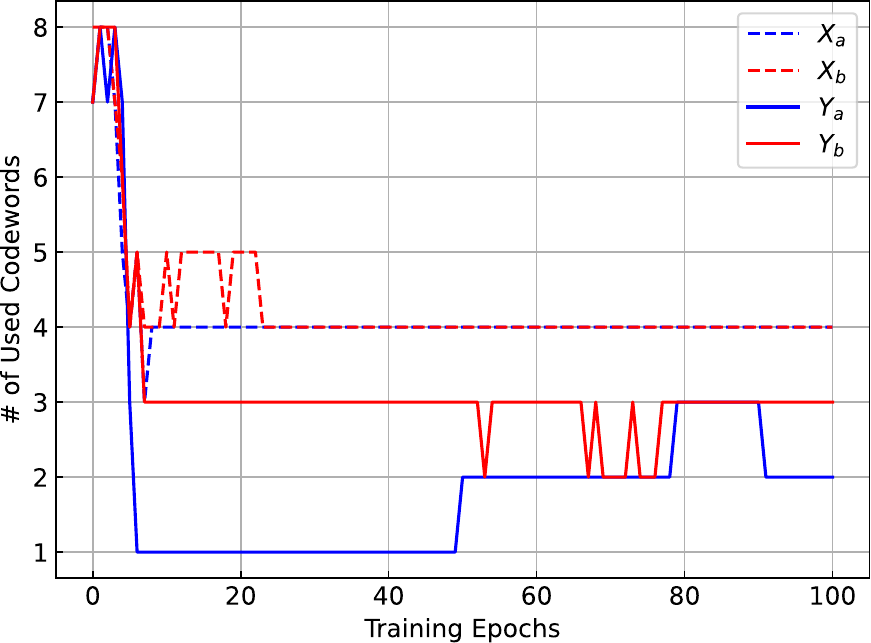}}
    {\caption{Level 1. Neural protocol: Causality-boosted neural protocol learning, yielding (a) average reward, (b) message entropy, and (c) message codeword sparsity variations during training.}}
    \label{fig:ECI}
\end{figure*}

After training completes, the feed-forward process of this actor NN is summarized as $s_j \to g\big (m_j^\text{up}, m_{\mathcal{J}\backslash j}^\text{up}\big) \to a_j$, where $g(\cdot)$ represents the NN layers at the BS. Comparing this with a single UE scenario associated with $s_1 \to g(m_1^\text{up}) \to a_1$, we can identify that one key causation is $\big(s_j,m_{\mathcal{J}\backslash j}^\text{up}\big) \to a_j$, characterizing the effectiveness of emergent messages in inter-agent coordination. With two UEs, this is recast as the following non-linear causal model:
\begin{align}
a_j &= f(s_j, m_{-j}^\text{up}) \label{Eq:SCM1}
\end{align}
where $f(\cdot)$ is a non-linear function, and the subscript $-j$ denotes the UE other than the $j$-th UE. According to entropic causal inference \cite{kocaoglu2017entropic}, spurious correlations, i.e., anti-causal models, can be ignored by reducing $H(m_j^\text{up})$ while satisfying the following entropic inequalities:
\begin{align}
H(m_{-j}^\text{up}) &\geq  H(U_j) \label{Eq:ECI1}
\end{align}
where $U_j = \prod_{i=1}^2 H(a_j|s_j=i)$ since the observations on $a_j$'s are conditionally independent.

To penalize the violation of these entropic inequalities while reducing $H(m_j^\text{up})$,
and $H(m_2^\text{up})$, we consider the reguliarzer~$\mathcal{L}_\text{EC}$ to be minimzed:
\begin{align}
 \max \!\Big(0,  H(U_1) \!-\!  H(m_2^\text{up}) \Big) \! +\! H(m_2^\text{up}) \!=\! \underbrace{\max\!\Big(H(U_1),  H(m_2^\text{up})  \Big)}_{:=\mathcal{L}_\text{EC}},
\end{align}
where  $\max(0,x-y)$ is a hinge regularizer, which penalizes if $x>y$ and otherwise we have $0$. 
Fig~\ref{fig:ECI}(a) shows that applying this entropic causal regularizer $\mathcal{L}_\text{EC}$ increases average reward, while applying $-\mathcal{L}_\text{EC}$ decreases the reward, corroborating the effectiveness of the proposed entropic causality boosting.

To calculate and reduce the entropy $H(m_j^\text{up})$, following \cite{mordatch2018emergence}, we consider a set of $8$ codewards, and sample each message from the codeword following to a Dirichlet distribution, i.e., $H(m_j^\text{up}) \sim \text{Dir}(\alpha)$ with the concentration parameter $\alpha>0$. Fig~\ref{fig:ECI}(b) shows that the entropy $H(m_j^\text{up})$ of messages (solid lines) is no smaller than $H(U_{-j})$, satisfying the entropic inequalities in \eqref{Eq:ECI1}. Furthermore, as observed in Fig~\ref{fig:ECI}(c), decreasing $H(m_j^\text{up})$ facilitates code sparsity in both uplink and downlink messages. For instance, $H(m_1^\text{up})$ decreases from $2.51$ to $1$, during which the number of codewords for $m_1^\text{up}$ is reduced from $8$ to $4$. Similarly, the number of codewords for $m_1^\text{dn}$ is reduced from $8$ to $3$. Such induced code sparsity can make Level 2 MAC more compact as will study in the next section.

\section{Level 2 MAC: Neural Network-Oriented Symbolic Protocol}

\subsection{Motivation}

\textbf{Limitation of Level 1 MAC}.\quad
While effective in achieving high performance, Level 1 MAC inherits the fundamental limitations of NNs, in terms of computing efficiency, management flexibility, and generalization. In essence, task-oriented neural protocols rely on NN architectures that are over-parameterized for ease of training~\cite{bubeck2021universal}, incurring high computation, communication, and memory costs at execution.
% ~\cite{aicompute}. 
The NN is also a black-box function, making the protocol operations uninterpretable. Furthermore, it requires re-training in dynamic environments, such as the number of users, channel statistics, and tasks, thus hindering immediate reconfiguration and reducing the flexibility in protocol operations. Lastly, even under the same environment, gradient-based training does not guarantee obtaining the NN with the same model parameters
% ~\cite{becker2020geometry}
, leading to protocol fragmentation.

\textbf{Machine Language and Semantic Entropy}.\quad
At an application layer level, language-based machine communication protocols have been studied, including the knowledge query and manipulation language (KQML) \cite{finin1994kqml}.
% and the agent communication language (ACL) \cite{labrou1999agent}. 
These methods first build the structure of invariant concepts and their relations as a knowledge base, and then feed actual data into the knowledge base to generate communication messages. Similarly, the current generation of the Web, i.e., Web 3.0 also known as the semantic web, employs the same idea to make Internet data machine-readable. It first creates an ontology graph, and feeds actual data into the ontology graph to produce a knowledge graph (KG) representing semantic relations in Web documents. For instance, consider an ontology graph \{\textsf{Paper} $\overset{\text{hasAuthor}}{\xrightarrow{\hspace*{20pt}}}$\textsf{Author}\} where  nodes and the edge represent concepts and their relations, respectively. Feeding real data into this ontology yields a KG including a single sentence, e.g.,  \{\cite{openai2023gpt4}$\overset{\text{hasAuthor}}{\xrightarrow{\hspace*{20pt}}}$\textsf{OpenAI}\}. This KG can be augmented by associating its truthfulness in the range of $0$ to $1$, which becomes compatible with the definition of semantic entropy that measures the randomness of sentences or propositions \cite{Bao11}. For instance, the proposition ``\cite{openai2023gpt4}$\overset{\text{hasAuthor}}{\xrightarrow{\hspace*{20pt}}}$\emph{OpenAI}" is associated with the truthfulness $1$, resulting in its semantic entropy $0$.

\subsection{Key Technique: Neuro-Symbolic Protocol Transformation}

To overcome the fundamental limitations of neural protocols in Level 1 MAC, Level 2 MAC aims to reveal protocol operations explicitly. In fact, the NN architecture of neural protocols is not always mandatory at execution after training. Instead, treating a trained NN as a simulator, it is possible to extract symbolic protocol operations (i.e., NN outputs) by feeding agents' possible states (i.e., NN inputs). In this neural-to-symbolic transformation, the NN's data feed-forwarding architectural order determines the causality in protocol operations, e.g., \{\textsf{State}$\to \textsf{Msg}^\text{up} \to \textsf{Msg}^\text{dn} \to$\textsf{Action}\}, where \textsf{State} and \textsf{Action} are the initial NN layer input and the final layer output while $\textsf{Msg}^\text{up}$ and $\textsf{Msg}^\text{dn}$ are hidden layer outputs. When feeding actual agents' states, the NN outputs represent a rule of certain protocol operations, e.g., \{\textsf{full buffer}$\to m^\text{up} \to m^\text{dn} \to$\textsf{transmit}\} with emergent signaling messages $m^\text{up}$ and $m^\text{dn}$. Such causal structure and operational rules of the protocol play the same roles as the ontology graphs and KGs in semantic language \cite{finin1994kqml}.

Due to the over-parameterized nature of NNs, even limited inputs may produce a number of outputs at different NN layers, resulting in many redundant protocol rules. To reduce this redundancy, on the one hand, trivial connections that do not change final action can be skipped. For instance, with \{$\textsf{state}_1\to m^\text{up}_1 \to m^\text{dn}_1 \to\textsf{action}_1$\} and \{$\textsf{state}_1\to m^\text{up}_2 \to m^\text{dn}_2 \to\textsf{action}_1$\}, we have \{$\textsf{state}_1\to\textsf{action}_1$\}, which corresponds to grant-free access in traditional MAC \cite{anand2020joint}. On the other hand, similar messages can be clustered to reduce the redundancy in protocol rules. Consequently, clustered messages $\bar{m}^\text{up}$ and $\bar{m}^\text{dn}$ become Level 2 MAC message symbols that are invariant to minor input-output variations.

However, clustering similar messages may involve combining their slightly different connections. For this reason, the same message does not always lead to consistent protocol rules. In natural language, humans can differentiate multiple meanings of a single word based on context. Inspired from this, a context value $p\in[0,1]$ is assigned to each connection ``$\to$," which quantifies the generation frequency of the connection during its original neural protocol operations. After finalizing this neural-to-symbolic transformation, the resultant set of protocol rules yields a symbolic protocol that emulates the original neural protocol without running any NNs. This transformation enables to make protocol operations reconfigurable, compute-efficient, and comparable with other protocols as we shall elaborate in the following subsections.

\subsection{Key Principles: Graph and Semantic Entropy} \label{Sec:Lv2principle}

The structural and operational complexities of a symbolic protocol can be quantified using graph entropy \cite{Liu22} and semantic entropy \cite{Bao11}, respectively. First, consider that a symbolic protocol is a graph $G=(V,E, A)$ where its vertices $V$ represent agent states, actions, and signaling messages, while the adjacency matrix $A$ identifies feasible operations or their occurrence frequency out of of all possible edge connections $E$. To measure the entropy of this graph, the von Neumann entropy $H_{\text{v}}(G)$ of $G$ has been widely adopted to measure the entropy of such graphs \cite{Liu22}, which is based on two matrices that encode the graph's topology: the adjacency matrix $A=[a_{ij}]$ with $(i,j)\in E$, and the graph Laplacian matrix $L= D-A$. Here, $D$ is a degree matrix $D = \text{diag}(d_{1}, \cdots, d_n)$, where $d_i$ denotes the degree of the vertex $i \in V$ with $|V|=n$. Consequently, $H_v(G)$ is defined as follows:
\begin{align}
H_v(G) = -\sum_{i=1}^n \frac{\lambda_i}{\text{tr}(L)  } \log_2 \left( \frac{\lambda_i}{\text{tr}(L)}\right) \leq  \log_2(n), \label{Eq:GraphEntropy}
\end{align}
where $\lambda_i$ is the $i$-th smallest eigenvalue of $L$, and $\text{tr}(L)=\sum_{i=1}^n \lambda_i$.
% It has been known that is $H_v(G)$ bounded as: 
% \begin{align}
%  -\sum_{i=1}^n \frac{d_i}{ \text{tr}(L) } \log_2 \left( \frac{d_i}{\text{tr}(L)}\right)  -c\frac{\text{tr}(A^2)}{\text{tr}(L)} \leq H_v(G)  \leq  \log_2(n),
% \end{align}
% where $c = {\log_2 e}/{\min_{i} \lambda_i}$. 
% These bounds confirm that $H_v(G)$ captures the topological complexity of an SPM $G$ only using the sizes and connections of $G$. 
Notably, the upper bound of $H_v(G) $ implicates that the maximum structural complexity of a symbolic graph is determined by the size $n$ that includes the number of message codewords. As studied in Sec.~\ref{Sec:Lv1Case}, the number of message codewords decreases with the message entropy, underscoring the impact of causality-boosted neural protocol learning in Level 1 MAC on reducing the structural complexity of Level~2 MAC protocols.

Next, a symbolic protocol can also be viewed as a set~$C$ of logic clauses. Each clause $c$ describes a unit protocol operation, which is structured as a single connection of two neighboring vocabularies $A$ and $B$ with its frequency context $p$, i.e., $A\overset{p}{\to} B$. This representation coincides with the form used in semantic entropy that measures the truthfulness of a text proposition \cite{Bao11}. 
% Specifically, we can treat the logical clause $c$ in symbolic protocol as a text proposition, and interpret the frequency context $p$ of $c$ as the truthfulness of $c$. 
Consequently, the semantic entropy $H_s(c)$ of $c$ is given as $H_s(c) = -p\log(p) + (1-p)\log(1-p)$ \cite{Bao11}. Extending this, we can calculate the average semantic entropy of all $|C|$ clauses \cite{Jinho_2022_SemCom}:
\begin{align}
    H_s(C)= \frac{1}{|C|}\sum_{c \in C} H(c). \label{Eq:SemanticEntropy}
\end{align}
This $H_s(C)$ quantifies the operational complexity of the symbolic protocol $C$. We will utilize this metric to compare different symbolic protocols in the next subsection.

\begin{figure}
    \centering
    \includegraphics[width=.85\columnwidth]{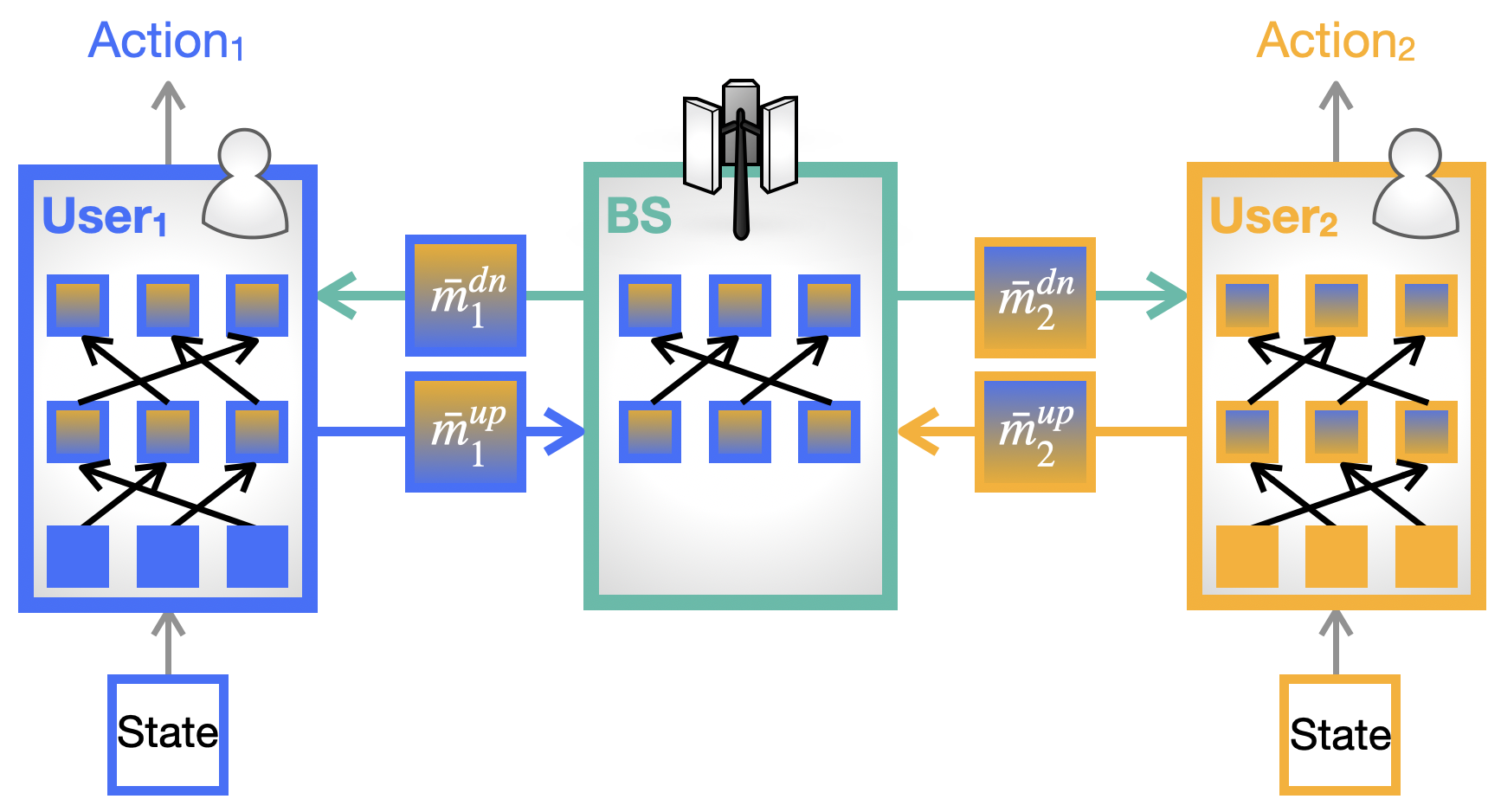}
    \caption{Level 2. Symbolic protocol: A single-cell network architecture example with two users.}
    \label{Fig:Lv2architecture}
\end{figure}

\subsection{Case Study: Symbolic Protocol Manipulation and Evaluation} \label{Sec:Lv2case}

\begin{figure}
    \centering
    \includegraphics[width=\columnwidth]{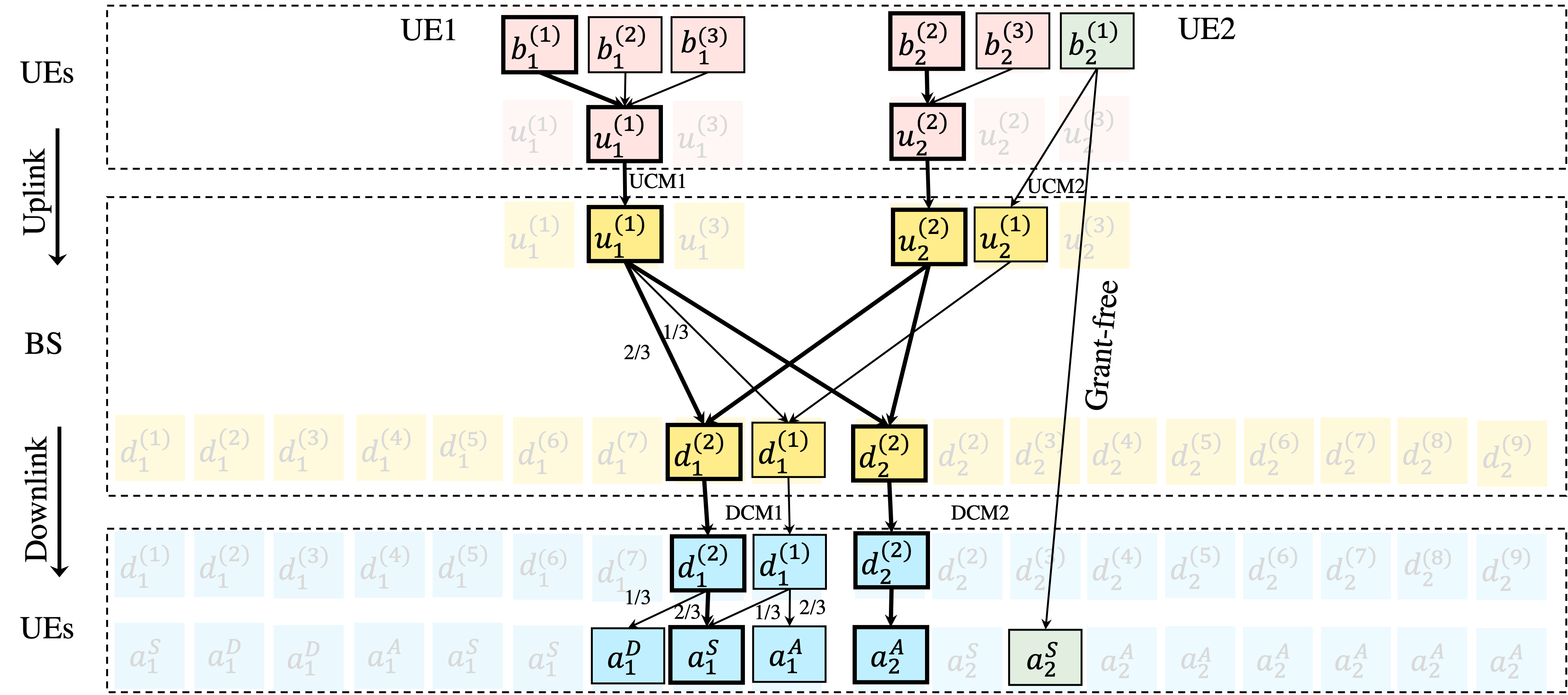}
    \caption{Level 2. Symbolic protocol: Extracted neural protocol before (transparent) and after symbol clustering (solid).}
    \label{Fig:Lv2case_extract}
\end{figure}

To showcase the transformation from Level 1 MAC to Level 2 MAC, as visualized in Fig.~\ref{Fig:Lv2architecture}, we revisit the two-user scenario in Sec.~\ref{Sec:Lv1Case} and its neural protocol in Level 1 MAC. Following \cite{seo2022towards}, the symbolic transformation of this neural protocol starts with extracting all possible inputs and outputs of the neural protocol. This extraction includes many redundant and similar protocol operations that can be skipped and merged. The resultant symbolic protocol is illustrated in Fig.~\ref{Fig:Lv2case_extract}. Compared to the original neural protocol, its transformed symbolic protocol accelerates protocol operations by 1,750x in terms of floating-point operations per second (FLOPS), and occupies only 0.02\% of the memory.

\begin{figure}
    \centering
    \subfigure[Symbolic manipulation]{\includegraphics[width=.49\columnwidth]{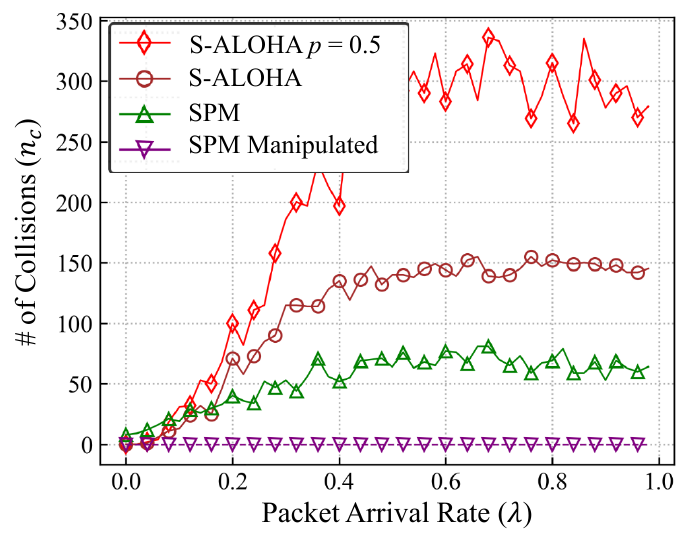}} 
    \subfigure[Semantic entropy based selection.]{\includegraphics[width=.49\columnwidth]{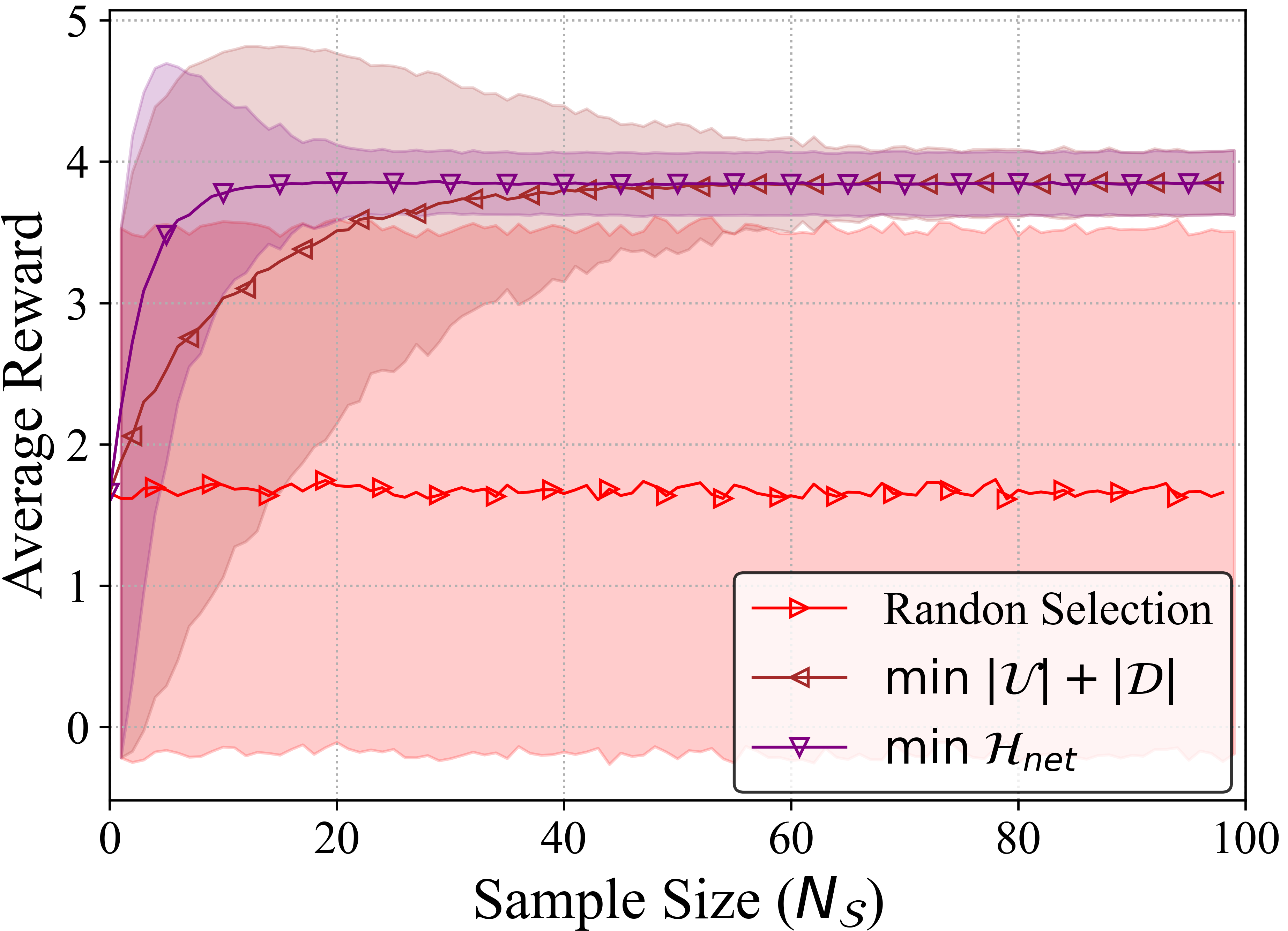}}
    \caption{Level 2. Symbolic protocol: (a) symbolic manipulation for collision avoidance and (b) semantic entropy for best symbolic model selection.}
    % \caption{Min Entropic SPM - SPM schematic illust, Selection reward}
    \label{Fig:Lv2case_result}
\end{figure}

Furthermore, symbolic protocols can be directly adjusted by changing unit protocol operations. To facilitate this reconfiguration, we can represent a symbolic protocol in form of the {probabilistic logic programming language (ProbLog)}~\cite{problog}, a logic programming language widely used in the field of symbolic artificial intelligence (AI).
% ~\cite{symbolicAI_problog}. 
Following the syntax of ProbLog, the logic clause $c$ ``$A\overset{p}{\to} B$" is written as $c = \langle p::B\text{ :- } A \rangle $. This ProbLog representation of the symbolic protocol facilitates flexible protocol manipulations by adding, removing, and connecting logic clauses. For instance, suppose another clause $c' = \langle p::B\text{ :- } A' \rangle $ that shares the outcome $B$ with $c$ but has a different cause $A'$. This case coincides with a collision event in MAC operations. As observed in Fig.~\ref{Fig:Lv2case_result}(a), in the neuro-symbolic semantic protocol, this collision can be instantly avoided by removing one of $c$ and $c'$, which is not feasible in its original neural protocol.

Lastly, as discussed in Sec.~\ref{Sec:Lv2principle}, the symbolic protocol's structural complexity can be measured using graph entropy $H_v(G)$ in \eqref{Eq:GraphEntropy}, which is upper bounded by the size~$n$ of the protocol, including state and action dimensions as well as the number of uplink and downlink message codewords. Recall that causality-boosted neural protocol learning in Sec.~\ref{Sec:Lv1Case} reduces the number of uplink and downlink message codewords from $8$ to $4$ and $3$, respectively. With $3$ states and $3$ actions, $n$ is reduced from $22$ to $13$. Therefore, we conclude that the use of causality-boosted neural protocol learning in Level 1 MAC achieves the 17.1\% reduction of the original $H_v(G)=4.46$ into $3.70$ in Level 2 MAC. On the other hand, semantic entropy quantifies the protocol's operational complexity. In practice, simpler or more definite protocol operations are preferable. Utilizing this, we can compare multiple protocols emerged from the same environment, and select the one achieving the lowest protocol entropy as the best symbolic protocol. Fig.~\ref{Fig:Lv2case_result}(b) demonstrates that this entropy based selection of protocols achieves higher average reward, compared to random selection and the selection based on the minimal numbers of uplink and downlink codewords.

% 3,8, 8, 3 =22, log2(22)=4.46
% 3, 4, 3, 3 = 13, log(13) = 3.70

\section{Level 3 MAC: Language-Oriented Semantic Protocol}

\subsection{Motivation}

\textbf{Limitation of Level 2 MAC}.\quad
Unlike Level 1 MAC protocols requiring re-training for any protocol adjustments, Level 2 MAC protocols permit direct manipulations. However, the scope of such adjustments should be confined within the protocol's original task knowledge. Violating this knowledge interpolation requirement may compromise the effectiveness of protocol operations in given tasks. One potential workaround is to leverage the memory-efficiency of Level 2 MAC. In other words, one can pre-train multiple neural protocols across various tasks, and utilize a portfolio of their symbolic protocols storing large task knowledge as discussed in \cite{seo2022towards}. Nevertheless, this strategy still hinges on task knowledge interpolation, and becomes challenging when it requires task knowledge extrapolation in as non-stationary task environments.

\textbf{Large Language Models and Cross-Modal Generative Models}.\quad    
As discussed above, merely enhancing task knowledge is not a sustainable strategy for coping with unseen task environments. Instead of taking this route, one can take the opposite direction by seeking general knowledge that can later be tailored to task-specific knowledge. The pursuit of this general knowledge has been a topic of debate in the past, but the rise of large language models (LLMs) has brought it to the forefront of recent research discussions. An LLM is a large NN trained on a vast corpus of human language data. For example, to capture general knowledge in form of human language, Llama 2 structured with $70$ billion model parameters was trained using $2$ trillion tokens \cite{touvron2023llama}. This amounts to $1.5$ trillion text words, which is approximately $350$x the size of Wikipedia that contains $4.3$ billion words. To cater to non-linguistic data modalities, LLMs can be integrated with cross-modal generative models, such as bootstrapping language-image pre-training (BLIP) \cite{li2022blip}, which translate images to texts and vice versa. Exploiting both LLMs and generative models enables human-language based protocol operations among machine agents and their language-based reasoning.

\subsection{Key Techniques: In-Context Learning and Linguistic Transformation}

Level 1 MAC and Level 2 MAC protocols only involve task-specific knowledge, and therefore lack the ability to extrapolate beyond their original task. To circumvent this limitation, Level 3 MAC employs human language as signaling messages, and seek to leverage the general knowledge embedded within LLMs. In this respect, Level 3 MAC is \emph{language-oriented}, and is intrinsically \emph{semantic protocols} where signaling messages mirror the semantics of human language. Employing human language imbues semantic protocols with inherent interpretability, ensuring native interoperability among machine agents and even with human operators. In addition, harnessing LLM's general knowledge offers flexibility in adjusting protocol operations, while enabling reasoning about it, making the underlying rationale of these adjustments transparent.

However, the general knowledge stored in LLMs can be overly broad to address specific tasks. Indeed, human language contains numerous polysemous words (i.e., words having multiple semantics), which poses potential misinterpretations during protocol operations. This emphasizes the need to refine general knowledge into task-specific knowledge. While LLMs are NNs that can be fine-tuned by adjusting their model parameters, their extensive parameter count incurs excessive computing resources and additional data. In-context learning is a viable alternative, which adjusts the LLM's outputs without altering its model parameters. While updating model parameters is about gaining more knowledge, in-context learning refines pre-trained LLM's general knowledge by delimiting the scope of interest within the LLM's general knowledge in a latent space. Drawing a parallel with humans who can discern intended semantics through additional examples and contexts, in-context learning demonstrates few task-related examples written in texts to the LLM, thereby deriving the LLM associated with task-conditional general knowledge that functions as task-specific knowledge.

While LLMs are rooted in human language, their general knowledge is not limited to linguistic data. Actually, the training data of LLMs encompasses a substantial amount of non-linguistic information, including text captions associated with videos and programming codes. Moreover, LLMs can readily interpret non-linguistic data given the right mappings between linguistic and non-linguistic data. In this regard, cross-modal generative models are instrumental. For instance, BLIP is a pre-trained NN that retains cross-attention information between images and their corresponding text captions, which facilitates conversions from images to text tokens (i.e., image-to-text, I2T) and the reverse (i.e., text-to-image, T2I).

\subsection{Key Principles: LLM Latent Space Theory and Prompt Engineering}

\textbf{LLM Latent Space Theory}.\quad
An LLM is capable of understanding the meaning or intention of a text prompt, which is crucial to enable decision-makings and reasoning over semantics. A recently proposed LLM's latent space theory explains how this works \cite{jiang2023latent}. Precisely, consider a text prompt $x$ is generated from an unknown latent intention $\theta_x \in q(\theta)$. The prompt $x$'s inherent ambiguity is quantified as $\epsilon(x)$, which satisfies: $\Pr(\theta_x|x) \geq 1-\epsilon(x)$. If an LLM is trained using an infinite number of data samples, the probability $p(y|x)$ of LLM's language generation $y$ given $x$ approximates the probability $q(y|x,\theta_x)$ of language generation with its unknown latent intention $\theta_x$, i.e.,
\begin{align}
\big| p(y|x) - q(y|x,\theta_x) \big| \leq \epsilon(x). \label{Eq:LLM1}
\end{align}
Here, the approximation error is bounded by $x$'s inherent ambiguity $\epsilon(x)$. 
This latent space theory demystifies how in-context learning works as follows. Consider a set $c_T =[ i_1, o_1, \cdots, i_T,o_T]$ of in-context learning samples, where each input-output pair $\{i_t,o_t\}$ is generated from a common intention $\theta_{c_T}$. With $c_T$, \eqref{Eq:LLM1} is recast as:
\begin{align}
\big| p(y|x, c_T, i_{T+1}) - q(y|i_{t+1},\theta_{c_T}) \big| \leq \epsilon(x)\epsilon(i_{T+1}) \prod_{t=1}^T \epsilon(i_t, o_t). \label{Eq:LLM2}
\end{align}
This implies that increasing $T$ or equivalently presenting more in-context learning samples improves LLM's language generation accuracy.

\textbf{Prompt Engineering}.
A proper prompt $x$ achieves the minimum inherent ambiguity $\epsilon(x) = 1-\Pr(\theta_x|x)$. To further reduce $\epsilon(x)$, it requires to increase $\Pr(\theta_x|x)$. To illustrate a viable solution, let us interpret $\Pr(\theta_x|x)$ as a T2I conversion task with the text prompt $x$ and its intended image~$\theta_x$. The accuracy of T2I conversion is often measured using the mutual information between the embeddings $Z_x$ and $Z_{\theta_x}$ of $x$ and $\theta_x$ in a common space, e.g., text tokens in BLIP. Recent studies heuristically showed that adding another embedding $Z_{x'}$ of $x'$ (e.g., image's skeleton, depth) generated from the same intention $\theta_x$ increases the mutual information, i.e., $I(Z_{\theta_x};Z_x|Z_{x'}) > I(Z_{\theta_x};Z_x)$ \cite{koh2023generating}. This result advocates $\Pr(\theta_x|x) < \Pr(\theta_x|x, x')$, reducing $\epsilon(x)$. The result underscores the need for utilizing different descriptions and various data modalities of the same semantics, as we shall study in the next subsection.

\begin{figure}
    \centering
    \includegraphics[width=.85\columnwidth]{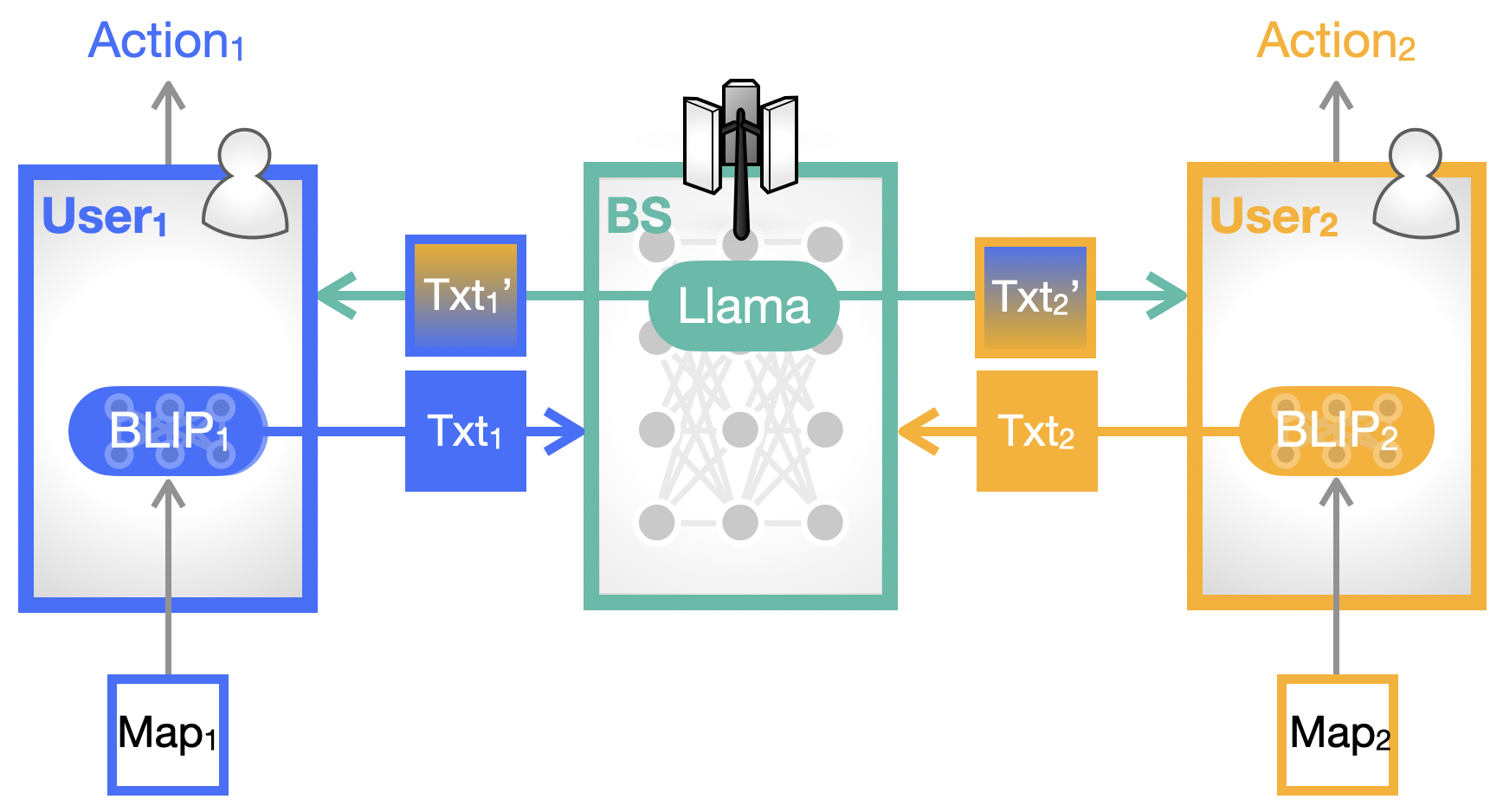}
    \caption{Level 3. Semantic protocol: A single-cell network architecture example with two users.}
    \label{Fig:Lv3architecture}
\end{figure}

\subsection{Case Study: Language-Oriented Multi-Agent Coordination}

\begin{figure}[t]
    \centering
    % \subfigure[Architecture with two agents.]{\includegraphics[width=.85\columnwidth]{Fig_Level3.png}}
    % \hspace{20pt}
    \subfigure[Global location map.]{\includegraphics[height=.32\columnwidth]{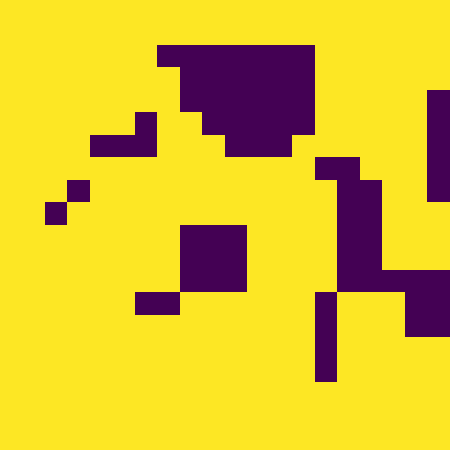}} \hspace{30pt}
    \subfigure[Global channel map.]{\includegraphics[height=.33\columnwidth]{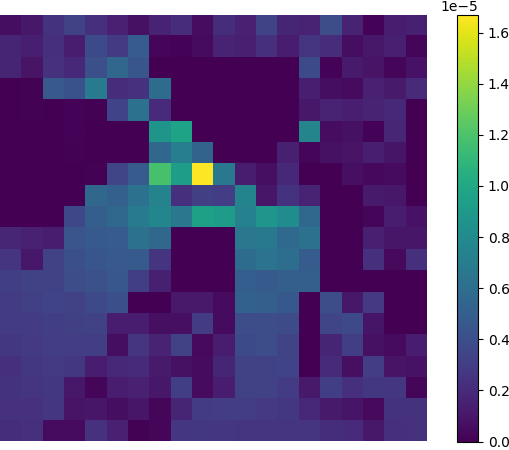}}
    \caption{Level 3. Semantic protocol: (a) architecture, (b) global location map, and (c) global channel map.}
    \label{Fig:Lv3Map}
\end{figure}

As a preliminary study to demonstrate the feasibility of language-oriented operations using Level 3 MAC, as visualized in Fig.~\ref{Fig:Lv3architecture}, we consider the single-cell network model studied in Sec.~\ref{Sec:Lv1Case} and \ref{Sec:Lv2case}, but with a remote navigation task with two agents in a two-dimensional grid. Given the unknown global location and channel maps as shown in Figs. \ref{Fig:Lv3Map}(a) and \ref{Fig:Lv3Map}(b), each agent aims to reach a pre-specified goal from an initial position while avoiding blockages and poor-channel areas that are visualized as dark areas in Figs.~\ref{Fig:Lv3Map}(a) and (b).

To this end, the $j$-th agent decides its next move to one of eight directions by observing 1) its local location map $s_j$ and 2) local channel map $h_j$ within eight directions, as well as downloading 3) a text message $m_j^\text{dn}$ from an BS. This downlink text message 3) is generated by an LLM stored at the BS after collecting uplink text messages $m_1^\text{up}$ and $m_2^\text{up}$ from both agents, i.e., $m_j^\text{dn}=g_\text{LLM}(m_1^\text{up},m_2^\text{up})$ where $g_\text{LLM}(\cdot)$ represents the LLM. The uplink text message $m_j^\text{up}$ is produced by a cross-modal generative model located at each agent, converting and merging 1) and 2) into a text message, i.e., $m_j^\text{up}=g_j(s_j,h_j)$ where $g_j(\cdot)$ represents the cross-modal generative model.

At the BS, we focus on the LLM process. The LLM receives $m^\text{up}=[m_1^\text{up}, m_2^\text{up}]$ from both agents, and produces their next move actions $m^\text{dn} = [m_1^\text{dn},m_2^\text{dn}]$ by asking a question $x$ ``which direction should Agent$_1$ and Agent$_2$ go?" This problem boils down to predicting $p(m^\text{dn}|x, m^\text{up})$. According to \eqref{Eq:LLM1}, the prediction error is bounded by the inherent error $\epsilon\big(x, m^\text{up}\big)$ of the messages. To reduce this error, following [REF], we add meta-instructions $x'$ on the problem and restrictions. Since $x'$, $x$, and $m^\text{up}$ are generated from the same task-specific intention $\theta$, adding $x'$ contributes to decreasing $\epsilon\big(x, m^\text{up}\big)$. Next, according to \eqref{Eq:LLM2}, in-context learning can further reduce the prediction error. To this end, we add $3$-shot examples $c_3=[i_1,o_1,i_2,o_2,o_3,o_3]$. The original question, meta-instructions, and one of $3$-shot examples are provided in Fig. \ref{Fig:Lv3prompt}.

\begin{figure}
    \centering
    \subfigure[Agent$_1$ trajectory.]{\includegraphics[width=.31\columnwidth]{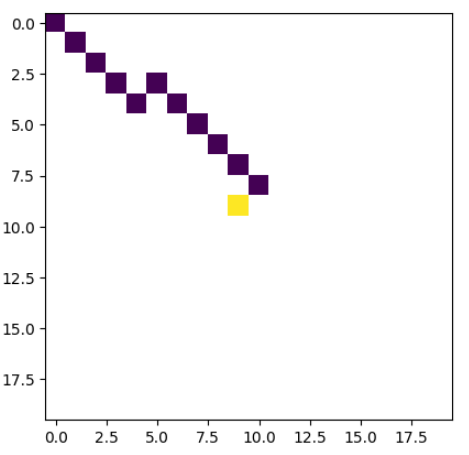}}
    \subfigure[Agent$_2$ trajectory.]{\includegraphics[width=.31\columnwidth]{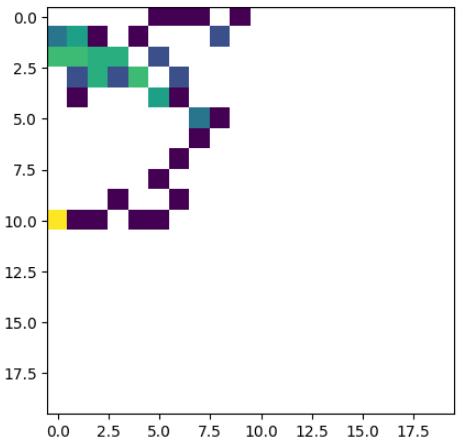}}
    \subfigure[Agent$_2$ trajectory, without. Agent$_1$.]{\includegraphics[width=.31\columnwidth]{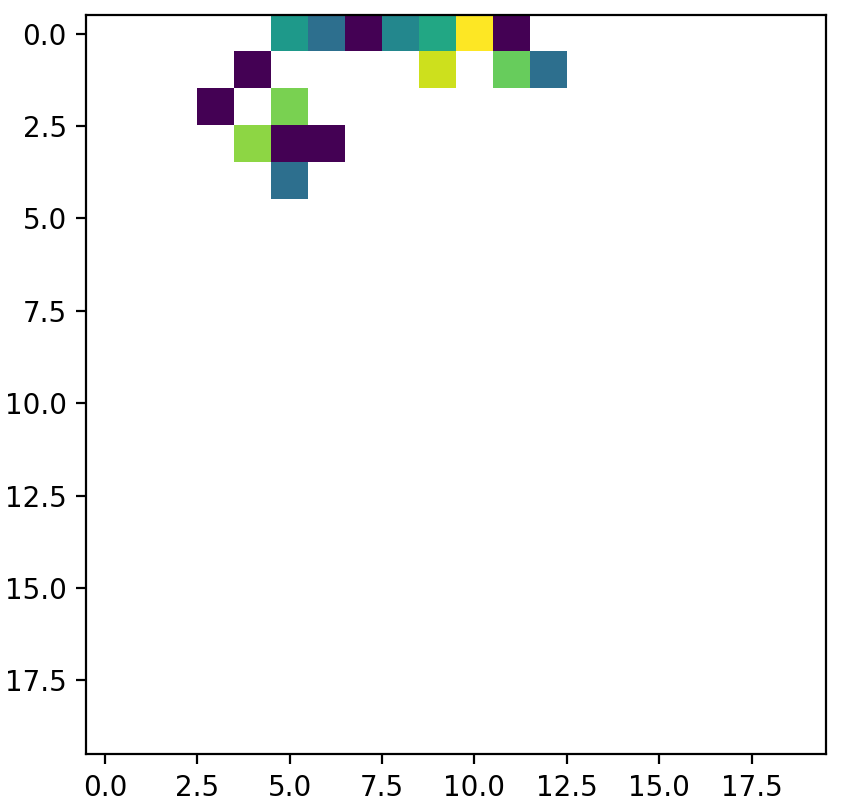}}
\caption{Level 3. Semantic protocol: Navigation trajectories of (a) Agent$_1$, (b) Agent$_2$, and (c) Agent$_2$ without Agent$_1$.}
\label{Fig:Lv3trajectory}
\end{figure}

Fig.~\ref{Fig:Lv3trajectory}(a) and (b) visualize the agent navigation trajectories upon running the aforementioned language-oriented semantic protocol. Precisely, Fig.~\ref{Fig:Lv3trajectory}(a) shows that Agent$_1$ successfully reaches its goal at $(9,9)$ from the initial position $(0,0)$. Likewise, Fig.~\ref{Fig:Lv3trajectory}(b) shows that Agent$_2$ successfully reaches at its goal at $(10,0)$ from the initial position $(0,9)$. Agent$_2$ is initially located in between the map's edge and a blockage as shown in Fig.~\ref{Fig:Lv3trajectory}(b). This challenging environment makes Agent$_2$ reach the destination slower than Agent$_1$. Remarkably, without navigating concurrently with Agent$_1$, Agent$_2$ fails to reach the goal as shown in Fig.~\ref{Fig:Lv3trajectory}(c), underscoring the effectiveness of the downlink message $m^\text{dn}$. This ablation study suggests that the LLM understands the history of Agent$_1$'s uplink messages $m_1^\text{dn}$ and its underlying location-channel environments, and is capable of exploiting this knowledge for guiding Agent$_2$ through $m^\text{dn}$. In these simulations, we used Llama 2 for the LLM and BLIP for the cross-modal generative model. Given a hand-crafted location map, the global channel map was constructed using the ray-tracing simulator \cite{choi2023withray}.

\begin{figure}[t]
\centering
\begin{myprompt}\footnotesize
% the instructions to the optimizer LLM that explain the optimization goal and instruct the model how to use the above information. The meta-instructions may also specify the desired generated instruction format for easier parsing.
\textbf{Meta-Instructions}.\quad
Your task is to guide Agent$_1$ and Agent$_2$ to their goals. 
Each step, you'll receive a message from both with directions to the goal, possible moves, and channel strength per direction.
Based on the given information, select one next direction from the eight options: northwest, north, northeast, west, east, southwest, south, and southeast.
This is a way of determining direction: 
\begin{enumerate}
    \item Decide among the directions the Agent can move.         
    \item Ensure the Agent approaches the goal.            
    \item Avoid the weak channel for the Agent.
    \item If there is a weak channel is near the goal, avoid that direction.
\end{enumerate}
% \vspace{3pt}
Answer in format [Answer: Agent$_1$: (direction); Agent$_2$: (direction); Rationale: (explain)]. Choose one direction each. Choose differently if the current position repeats.

\vspace{3pt}
\textbf{Few-Shot Example}.\quad Question:     
Agent$_1$ said ``I can go to: north, south, east. 
            From the location ($x=5$, $y=6$) aiming for ($10$, $6$), the direction that goes closest to the goal is south. 
            The channel power is medium at east, north.
            The channel power is weak at south, so the communication service is unstable."
 Agent$_2$ said ``I can go to: northwest, west, southwest, east. 
            From the location ($x=6$, $y=8$) aiming for ($1$, $9$), the direction that goes closest to the goal is northeast. 
            Other directions like north and east will also get me closer. 
            The channel power is strong at northwest, west, southwest.  
            The channel power is medium at east."
            Which direction should Agent$_1$ and Agent$_2$ go? 
            [Answer: Agent$_1$: east; Agent$_2$: east; 
            Rationale: Agent$_1$ should choose east as it has medium channel power. The agent should avoid weak channels as much as possible, although south is a closer direction.
            Agent$_2$ should choose ``east" as it brings them closer to the goal and has medium channel power.
            Although other directions have strong channels, it is also important to arrive at the goal as soon as possible.]

\vspace{3pt}
\hrule
\vspace{3pt}

% \begin{wrapfigure}{r}{3pt}
%   \begin{tikzpicture}
% \node[alice,mirrored, minimum size=3pt];
% \end{tikzpicture}
% \end{wrapfigure}      
% \begin{wrapfigure}{l}{3pt}
%   \begin{tikzpicture}
% \node[bob,minimum size=3pt];
% \end{tikzpicture}
% \end{wrapfigure}      

\textbf{Question}.\quad
{Agent$_1$} said ``I can go to: south, east, southeast. From the location ($x=0$, $y=0$) aiming for ($x=9$, $y=9$), the direction that goes closest to the goal is southeast. Other directions like northeast, east, southwest and south will also get me closer. The channel power is medium at south, east, southeast."
{Agent$_2$} said ``I can go to: west, southwest, south, east, southeast. From the location ($x=0$, $y=9$) aiming for ($x=10$, $y=0$), the direction that goes closest to the goal is southwest. Other directions like northwest, west, south and southeast will also get me closer. The channel power is medium at west, south, east, southeast. The channel power is weak at southwest, so the communication service is unstable." Which direction should Agent$_1$ and Agent$_2$ go?

\vspace{3pt}
\textbf{Answer}.\quad Agent$_1$: southeast; Agent$_2$: west; Rationale: Agent$_1$ should choose ``southeast" as it brings them closer to the goal and has medium channel power. Agent$_2$ should choose ``west" as it brings them closer to the goal, avoiding ``southwest" due to weak channel power. 
\end{myprompt}
\caption{Language-oriented semantic protocol: LLM text prompts of meta-instructions, few-shot examples for in-context learning, and a question/answer.} \label{Fig:Lv3prompt}
\end{figure}

\section{Conclusion and Future Directions}

In this article we have categorized data-driven MAC protocols into three levels: Level 1. task-oriented neural protocol, Level 2. NN-oriented symbolic protocol, and Level 3. language-oriented semantic protocol. To grasp the opportunities and challenges presented by these three-level protocols, we delved into their underlying key techniques and principles. Selected case studies showed that information-theoretic methods hold immense potential for enhancing both Level~1 and Level~2 MAC protocols, and they also provide tools for evaluating the structural and operational complexities of Level~2 MAC.

To address severe task variability that cannot be addressed by the task-specific knowledge of Level~1 and Level 2 MAC, Level 3 MAC additionally utilizes generic knowledge embedded in LLMs. However, existing hardware and LLM architectures are not yet to achieve best efficiency, leading to considerable computation overhead and latency during execution. For instance, Llama incurs 13 GFLOPS of computation complexity and achieves 30 token (approx. 20 words) generation rate per second on consumer-grade GPUs. This contrasts sharply with the 8 FLOPS and 14K FLOPS observed in our simulations for Level 1 and Level 2 MAC, respectively. This discrepancy presents a trade-off between Level 2 and Level 3 MAC. Balancing this trade-off could be an interesting topic for future study. Furthermore, by extending the LLM latent space theory, it could be promising to develop novel information theoretic methods for optimizing Level 3 MAC. Lastly, considering that semantic communication has been extensively studied at the PHY layer, integrating semantic PHY and semantic MAC protocols would emerge as another intriguing topic for future exploration.

\section*{Acknowledgement}
The authors are grateful to Sejin Seo at Yonsei University, as well as Yongjun Kim and Junil Choi at KAIST, for their contributions to simulations and discussions. This work was supported in part by the Institute of Information \& communications Technology Planning \& Evaluation (IITP) grant (No. 2021-0-00347) funded by the Ministry of Science and ICT (MSIT), and in part by the Information Technology Research Center (ITRC) support program (IITP-2023-RS-2023-00259991) supervised by IITP. J. Park is the corresponding author (email: jihong.park@deakin.edu.au).

\bibliographystyle{ieeetr}
\bibliography{main.bbl}

\end{document}